# Climate Change: The Sun’s Role

***Gerald E. Marsh***
Argonne National Laboratory (Ret)
5433 East View Park
Chicago, IL 60615

E-mail: geraldemarsh63@yahoo.com

**Abstract.** The sun’s role in the earth’s recent warming remains controversial even though there is a good deal of evidence to support the thesis that solar variations are a very significant factor in driving climate change both currently and in the past. This précis lays out the background and data needed to understand the basic scientific argument behind the contention that variations in solar output have a significant impact on current changes in climate. It also offers a simple, phenomenological approach for estimating the actual—as opposed to model dependent—magnitude of the sun’s influence on climate.

**Introduction**

The role of the sun in climate change remains contentious not so much because there is no data to support the thesis, but because there is no accepted mechanism to explain the positive feedback that must underlie the sometimes dramatic changes in climate resulting from relatively small changes in solar irradiance or insolation. Changes in the latter, due to small quasi-periodic variations in the earth's orbital parameters—the Milankovitch theory, are thought to be responsible for past glaciations. The difficulty some have expressed about this theory is that the climate swings associated with the ice ages and inter-glacials seem to be too large for the small changes in insolation due to the cycles without postulating a very strong positive feedback mechanism. Another problem is that for the last million years or so the ice ages have a 100ky period. Previous to this the period was about 40ky. This does not fit well with the Milankovitch theory. A warming climate also appears to sometimes precede by about 10ky the change in insolation that supposedly was its cause. One possible reason for these inconsistencies is the possibility of additional periodicities in solar output. A summary of these inconsistencies and a possible explanation has been given by Ehrlich [1]. It should be noted, however, that there is a broad consensus supporting the Milankovitch theory.

When it comes to changes in irradiance, the small observed variations are thought to require a positive feedback of significant magnitude if they are to have a readily observable effect on climate [2]. A number of mechanisms for such amplification have been proposed in the literature [3]. One possibility is the heating of the stratosphere by solar ultraviolet radiation, which varies by several percent over the eleven year solar cycle compared to ~0.1% for radiation at the peak of the spectrum, followed by a dynamic coupling of the ultraviolet-heated stratosphere to the lower atmosphere. A second involves the effect of galactic cosmic rays on climate. The concentration here will be on the latter and in particular on the correlation between solar activity and variations in the earth's albedo.

The literature dealing with the connection between climate and solar variations is vast and goes back many, many years. This exposition does not pretend to do justice to this history, but is being written to a large extent in response to the 2007 Intergovernmental Panel on Climate Change (IPCC) *Summary for Policymakers,* which—despite much evidence to the contrary—greatly downplays the role of the sun on climate change since the IPCC's reference point of 1750. (The SPM is available at the IPCC website: http://www.ipcc.ch.)

Data showing a connection between solar variations and climate are often dismissed as mere correlations since there is no generally accepted theoretical basis to explain these correlations. This is hardly an acceptable position given that the strongest argument in favor of carbon dioxide driven climate change is the correlation between the rise in the concentration of carbon dioxide in the atmosphere, mostly due to human activity, and the warming of the climate especially since the 1970s. It should be remembered that, before the 1970s—and despite the relatively large rise in the concentration of carbon dioxide before 1970—most climatologists thought we were sliding into another ice age [4]. In the case of carbon dioxide, however, there is a generally accepted mechanism for linking changes in climate to variations in the concentration of this gas since it is an absorber of long-wavelength infrared radiation, and simple *1*-dimensional radiative equilibrium models suffice to illustrate the connection.

For those not already familiar with it, the key concept that will be used in this discussion is *radiative forcing*. It is defined as the change in net downward radiative flux at the tropopause resulting from any process that acts to perturb the climate system; it is measured in $w/m^2$. This definition is believed to be the most appropriate one to characterize the response of the earth's surface and troposphere to a radiative perturbation *after* the stratosphere has come to thermal equilibrium. It is based on the fact that the response time of the stratosphere to a radiative perturbation is much shorter than that of earth's surface and closely coupled troposphere. Once the stratosphere has come to thermal equilibrium, it can be argued that the change in flux at the tropopause and at the top of the atmosphere are essentially the same. Examples of radiative forcing are variations in the amount of solar radiation reaching the earth and changes in the concentrations of infrared-absorbing gases in the atmosphere.

Consider the case of carbon dioxide. Carbon dioxide is often labeled "the most important greenhouse gas" by the media and in popular literature. Those familiar with the field know that this is not the case. Water vapor (by far the most important greenhouse gas) is excluded because it is treated as a feedback in climate models. Unfortunately, this is not understood by most of the general public, nor by many policymakers for that matter. Carbon dioxide is a minor greenhouse gas responsible for only a small fraction of the earth's greenhouse effect [5]. Increasing the amount of carbon dioxide in the atmosphere by, for example, a factor of two does not double the amount of infrared radiation absorbed by this trace gas. The reason for this has to do with where the carbon dioxide absorption bands are located relative to the earth's emission spectrum, and the amount of carbon dioxide already in the atmosphere. Carbon dioxide has three absorption bands at wavelengths of 4.26, 7.52, and 14.99 micrometers (microns). The earth's emission spectrum, treated as a black body (no atmospheric absorption), peaks at between 15 and 20 microns, and falls off rapidly with decreasing wavelength. As a result, the carbon dioxide absorption bands at 4.26 and 7.52 microns absorb negligible amounts of thermal radiation compared to the band at 14.99 microns.

Despite its being only a trace gas in the earth's atmosphere, natural concentrations of carbon dioxide are great enough that the atmosphere is opaque even over short distances in the center of the 14.99 micron band. As a result, at this wavelength, the radiation reaching the tropopause from above and below the tropopause is such that the net flux is close to zero.

If this were the whole story, adding more carbon dioxide to the atmosphere would contribute nothing to the greenhouse effect and consequently could not cause a rise in the earth's temperature. However, even though the 14.99 micron band is essentially saturated, adding additional carbon dioxide does have an influence at the edges of the band. Because of this marginal effect, the change in forcing due to a change in carbon dioxide concentration is proportional to the natural logarithm of the fractional change in concentration of this gas. Specifically, the IPCC gives

$$\Delta F = \alpha \ln\left(\frac{C}{C_0}\right),$$

where $C$ is the concentration of carbon dioxide at the time of interest, $C_0$ is the concentration at a given reference time, and $\alpha$—having the dimensions of w/m$^2$—is the sensitivity of the climate to changes in carbon dioxide concentration. The values of $\alpha$ given by the IPCC "implicitly include the radiative effects of global mean cloud cover" [6].

This approximation breaks down for very low concentrations and for concentrations greater than 1000 ppmv, but is valid in the range of practical interest. The earth's temperature is therefore relatively insensitive to changes in carbon dioxide concentrations. Unlike the logarithmic dependence of the forcing, which is based on well understood physics, the coefficient $\alpha$, because it includes the effects of clouds—a major uncertainty in climate models—can be expected to be associated with large errors. And this is indeed the case.

The value of $\alpha$ used by the IPCC in their 1990 assessment was 6.3 w/m$^2$. In 2001 it was 5.35 w/m$^2$, and in the 2007 report it has climbed somewhat to 5.5 w/m$^2$. The last two values compared to the first imply an error range of 15-19%. As a result, given the uncertainties associated with $\alpha$, one might well say that the strongest argument for the anthropogenic forcing of climate through carbon dioxide emissions is also simply a correlation. Indeed, "… scientists began studying climate change based primarily on the observation that $CO_2$ was rising steeply and the notion, based on simple radiative-forcing arguments, that the atmospheric $CO_2$ burden could not continue to rise without producing some effect on climate." [3] The question really is, "How much?".

**The radiative balance of the Earth**

The sun is a main-sequence star whose luminosity in the past is given for $t \leq t_0$ by the well-founded expression [7]

$$L(t) = \left[1 + \frac{2}{5}\left(1 - \frac{t}{t_0}\right)\right]^{-1} L_0,$$

where $L_0$ is the current luminosity of the sun and $t_0$ is the present age of the sun. 4.6 billion years ago, when the sun first began to shine, $t/t_0$ was zero and today $t/t_0 = 1$. What this expression tells us is that, after its formation some 4 billion years ago, the temperature of the earth's surface would have been below the freezing point of water for the first 2 billion years of its existence [8]. But this is impossible since there are sedimentary rocks, which had to form in liquid water, that date back some 3.8 billion years. Indeed, the early earth was warmer than today since there was no glaciation prior to about 2.7 billion years ago. Whether this was due to a lower albedo or an enhanced greenhouse effect is not really known, although the evidence seems to point to the latter.

The most likely mechanism for warming the early earth is the carbon dioxide geochemical cycle where without liquid water the weathering of silicate rocks would cease, and carbon dioxide released by volcanoes would accumulate in the atmosphere until the temperature was above freezing thereby allowing weathering to resume. Assuming the silicate rocks can be represented by the mineral wollastonite ($CaSiO_3$), the chemistry behind the weathering process is as follows:

$$CaSiO_3 + 2\,CO_2 + H_2O \rightarrow Ca^{++} + 2\,HCO_3^{-} + SiO_2$$

$$Ca^{++} + 2\,HCO_3^{-} \rightarrow CaCO_3 + CO_2 + H_2O$$

The net result being,

$$CaSiO_3 + CO_2 \rightarrow CaCO_3 + SiO_2.$$

The importance of these observations in the present context is that while it has been argued above that carbon dioxide is a minor greenhouse gas, it is also true that very large changes in its concentration can significantly affect climate. The above resolution of the faint young sun problem shows that this was likely the case in the distant past.

Today, the temperature of the earth would also be below the freezing point of water were it not for the greenhouse effect. To show this one uses the Stefan-Boltzmann relation to determine the equilibrium temperature of the earth in the absence of the greenhouse effect. This involves some geometry. As seen from space, the earth looks like a disk of area $\pi r_e^2$, where $r_e$ is the radius of the earth. So, when averaged over the whole earth, which has a surface area of $4\pi r_e^2$, the amount of radiation per square meter reaching the earth from the sun must be multiplied by the geometrical factor $\pi r_e^2/4\pi r_e^2 = 1/4$. Because the earth also reflects some of the radiation back into space, we must also multiply by the factor $(1 - A)$, where $A$ is the albedo of the earth, equal on average to about 0.3. The appropriate Stefan-Boltzmann relation for radiative equilibrium is then

$$\frac{(1 - A)}{4} I = \sigma T_e^4,$$

where $T_e$ is the radiative equilibrium temperature of the earth, $I$ = 1365 $w/m^2$ is the solar irradiance at the earth's distance from the sun, and $\sigma = 5.67 \times 10^{-8}$ $w\ m^{-2}K^{-4}$ is the Stefan-Boltzmann constant. Solving for the temperature of the earth with these values of the constants gives $T_e$ = 255 $^oK$, well below the freezing point of water at 273 $^oK$. It is the greenhouse effect of the earth's atmosphere, mostly due to water vapor and clouds followed by trace gases like carbon dioxide, that cause the surface of the earth to have a temperature of some 33 $^oK$ higher. The radiative equilibrium temperature of 255 $^oK$ is still the radiating temperature of the earth into space—it must be if the earth is to be in equilibrium with the energy absorbed from the sun—but now it is the effective radiating temperature at the top of the atmosphere.

In the above energy-balance equation, emissivity of the earth was deliberately not included. This is because the emissivity of the earth's surface is generally very close to unity. Often one uses the "effective emissivity" to take into account the earth's atmosphere and clouds. In this way the greenhouse effect is included in the emissivity and the resulting equilibrium temperature is again 288 $^oK$.

In what follows, it will be necessary to relate changes in cloud cover to changes in radiative forcing. Unfortunately, this is not easily done. Nonetheless, a simple approximation can be obtained from the overall albedo of the earth. With an albedo of 0.3, the earth reflects 409.5 w/m$^2$. This corresponds to a radiative forcing of 102.5 $w/m^2$; since, according to the IPCC, clouds are responsible for up to two thirds of this (estimates range from 50-66%, so a value of 60% will be used here), their share is about 61.5 $w/m^2$. Assuming the mix of high- and low-level clouds remains the same, a 1% decrease in total cloud cover corresponds to a net forcing of 0.62 $w/m^2$. If the generally accepted overall climate sensitivity of 0.53 $^o$Cw$^{-1}$m$^2$ is used (this will be revisited later in the penultimate section of this paper), this forcing corresponds to a rise in global temperature of 0.33 $^o$C. If the decrease in cloud cover is related to an increase in solar activity—a connection that will be made later—the increase in temperature due to the direct increase in solar irradiance should be added to this value. In summary, a 1% decrease in global cloud cover—using the generally accepted overall climate sensitivity—corresponds to a global increase in temperature of at least 0.33 $^o$C. This value is well within the range of values found in the literature derived by far more intricate means [2].

There are two approaches to understanding changes in climate in response to variations in solar output: one can rely on climate models with their inherent errors due to the current limitations of our knowledge and ability to model the earth's non-linear response to such variations, or one can use a phenomenological approach. Given that it is generally admitted that our understanding of solar effects on climate—which may involve several different types of forcing—is limited, the second approach seems more attractive, at least until the interaction mechanisms are better understood.

## Solar variations and temperature

The actual irradiance from the Sun has been monitored by spacecraft since 1978, covering thus far only two of the Sun's eleven year cycles. During that period, the solar output varied over a range of 0.15%. Between the twelfth century Mediaeval Maximum and the Maunder Minimum of 1645-1715 (the time of the Little Ice Age) the brightness of the Sun is estimated to have decreased by as much as 0.5%, although recent work suggests this may be too high. Solar-type stars have also been found to have variations of 0.1% to 0.4%. These values seem innocuous, but in fact they can have a disproportionately large impact on climate in the presence of a strong positive feedback mechanism.

There is also evidence that the Sun has variations in output with periodicities of ~70-90 years, ~200 years, and ~2500 years. These solar variations may enter the climate system by affecting the Quasi-biennial Oscillation and the El Niño Southern Oscillation. When the historical record of El Niño events is compared to the record of sunspot numbers, El Niño events are found to be two to three times more frequent when sunspot activity and solar irradiance are low—as during the Maunder minimum.

Total solar irradiance in the past is difficult to determine and may—for the period since the Maunder Minimum (associated with the Little Ice Age) of the mid-1600s to the early 1700s—have an uncertainty of anywhere from 1 to 15 $w/m^2$. However, what really matters for the purposes of this paper is not the absolute uncertainty but the relative changes and uncertainties over the time period of interest. The best data available in 2001 are shown in Figure 1 below. Since that time, the data have been updated in 2004 and again in 2005; there has also been additional work on estimating the change in total irradiance by others in 2007. More will be said about this later. Note the correlation between sunspot number and total solar irradiance.

In case the figure below is printed and does not appear in color, the solid lower curve is red; the topmost curve is sunspot number in grey; and the middle curve is black.

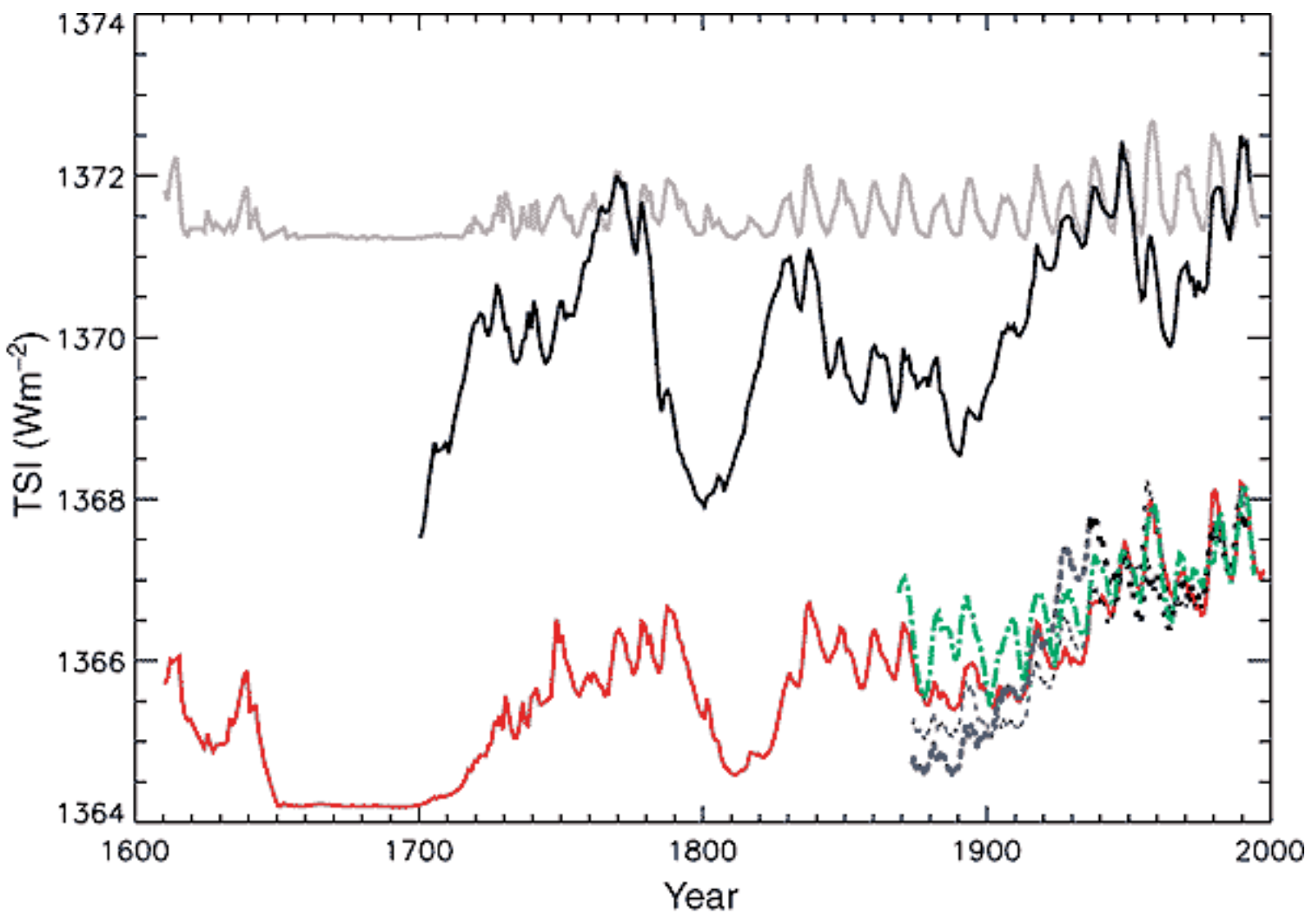

Figure 1: Reconstructions of total solar irradiance (TSI) by Lean et al. (1995, solid red curve), Hoyt and Schatten (1993, data updated by the authors to 1999, solid black curve), Solanki and Fligge (1998, dotted blue curves), and Lockwood and Stamper (1999, heavy dash-dot green curve); the grey curve shows group sunspot numbers (Hoyt and Schatten, 1998) scaled to Nimbus-7 observations for 1979 to 1993. [Fig. 6.5 and caption from *Climate Change 2001: The Scientific Basis*]

The very strong correlation between temperature change and changes in solar activity can be seen from work now over fifteen years old. Instead of the usual smoothed sunspot number, Friis-Christensen and Lassen [9] introduced solar cycle length as a new measure of solar activity, and compared the solar cycle length with observed temperature anomalies since 1860 relative to the period of 1951-1980. Note that the eleven year solar cycle actually has lengths that generally fall between nine and fourteen years. The result is shown in Figure 2. The correlation is striking. Note also that the decrease from 1945 to 1970—a time of rising carbon dioxide concentration.

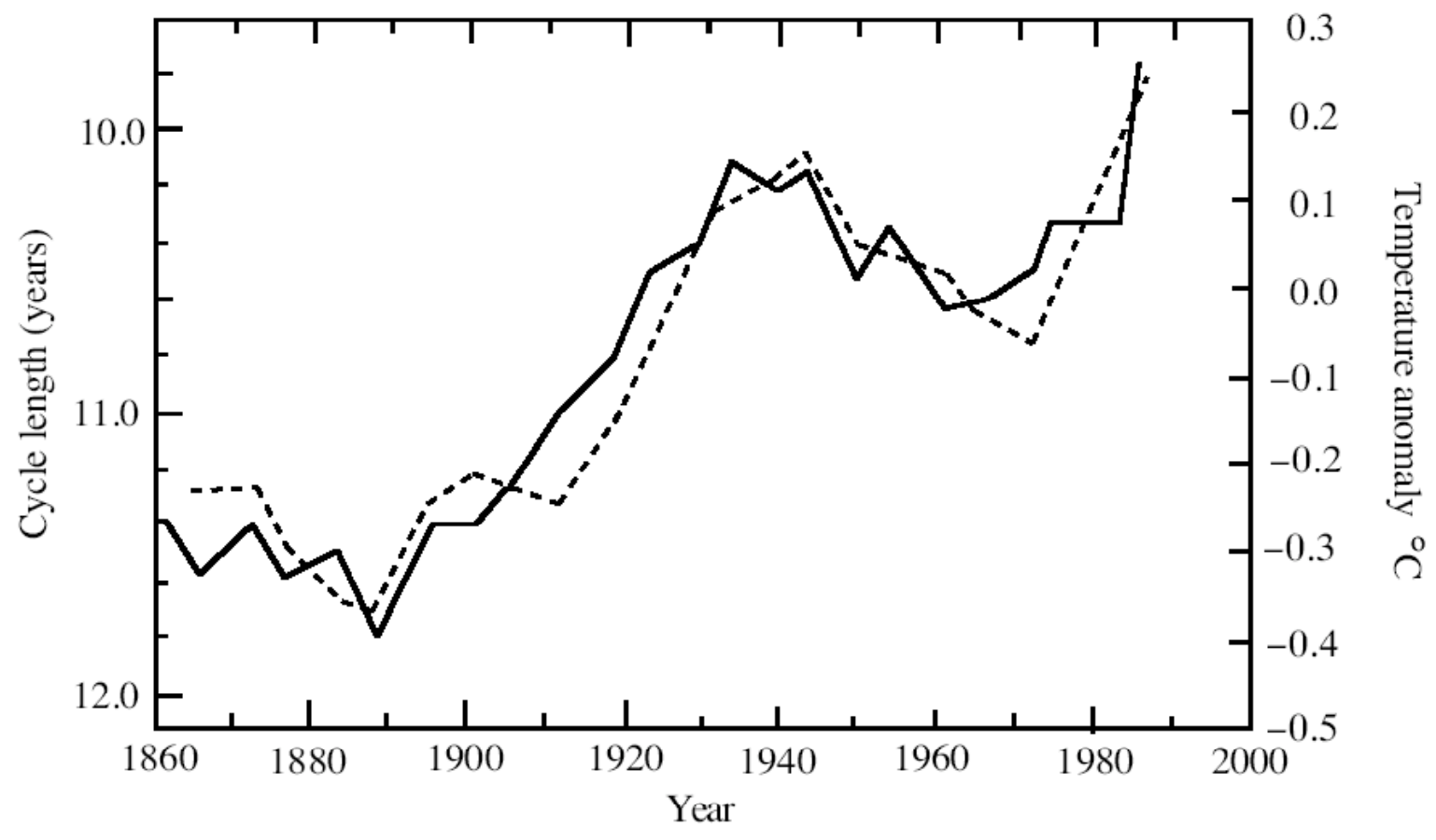

Figure 2. Solar cycle length (solid curve) and Northern Hemisphere temperature anomalies (dashed curve) as a function of year. The cycle length is plotted at the central time of the actual cycle. Note the inverse scale for solar cycle length: short solar cycles correspond to periods of high solar activity, and long cycles to reduced activity. [Adapted from E. Friis-Christensen and K. Lassen (1991)]

Since this work by Christensen and Lassen was completed, Christensen and Svensmark compared temperature and solar activity data going back to the second half of the sixteenth century [10]. Prior to 1750, they used solar cycle lengths based on auroral observations. The excellent correlation between solar activity and Northern Hemisphere land temperatures continues to hold as can be seen in Figure 3.

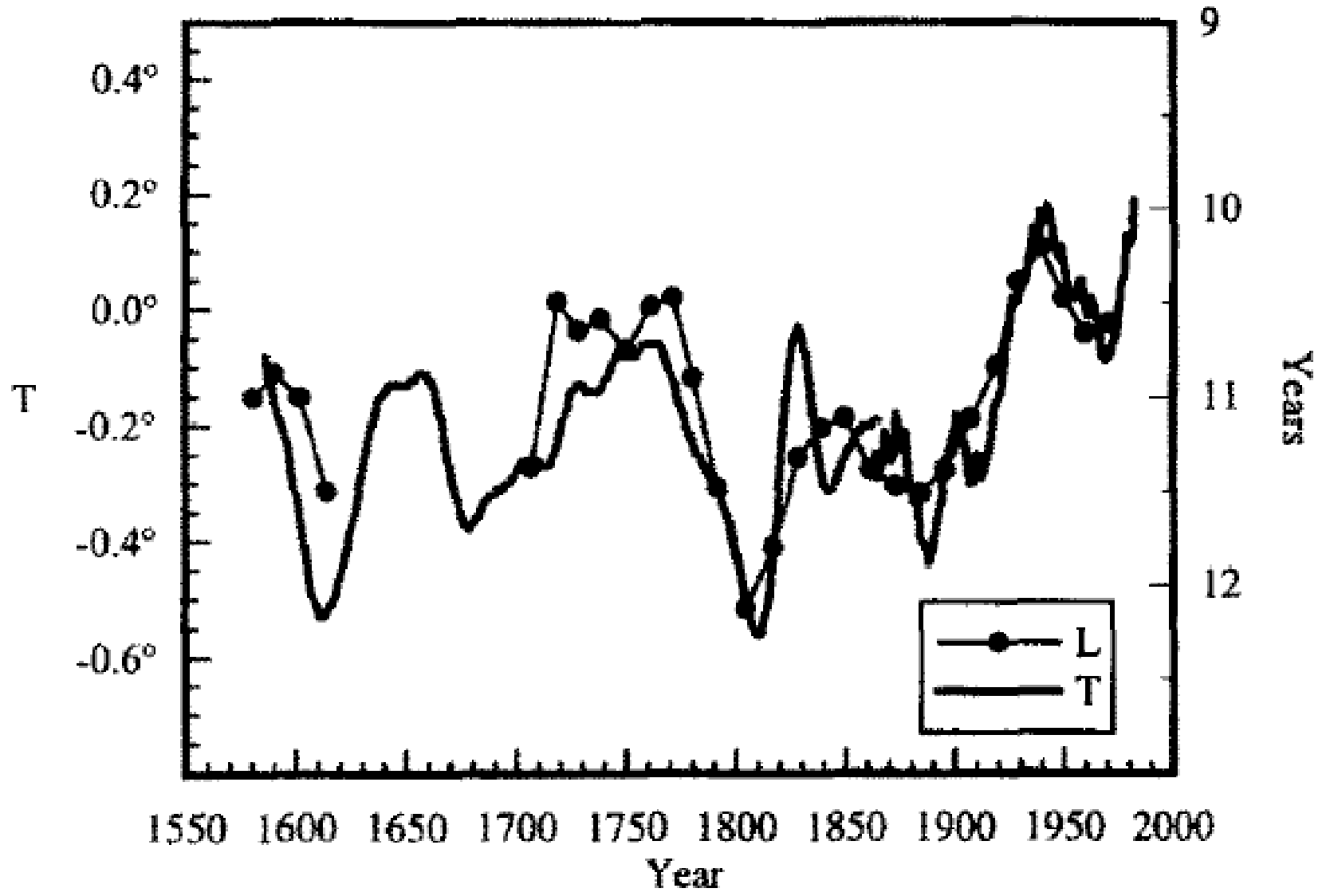

Figure 3. 11-year average values of the Northern Hemisphere land temperature (T), before 1860 estimated by means of tree ring analyses, and long term variation of solar activity expressed by the length (L, Years) of the sun spot cycle (before 1850 estimated by means of auroral observations). From Eigil Friis-Christensen and Henrik Svensmark, *Adv. Space Res.* Vol. 20, No. 415, pp. 913-921 (1997).

The correlations shown in Figures 2 and 3 are striking, to say the least. But what of the relatively large warming at the end of the twentieth century. Has the sun shown unusual activity during this period? The answer to this question is shown below [11]. As is readily seen, the high level of activity during the last sixty years transcends anything seen during the last 1150 years!

Notice in Figure 4 that the variation in $^{14}C$ has an inverted scale. The reason for this will become apparent below.

Intuitively, one might think that an increase in sunspots, since they appear dark, would reduce solar output—but the reverse is observed. This is because there is an increase in the brightness of regions of the sun known as faculae, which are associated with sunspots. The increase in the brightness of these faculae is the dominant factor in changing solar output rather than sunspot darkening [12].

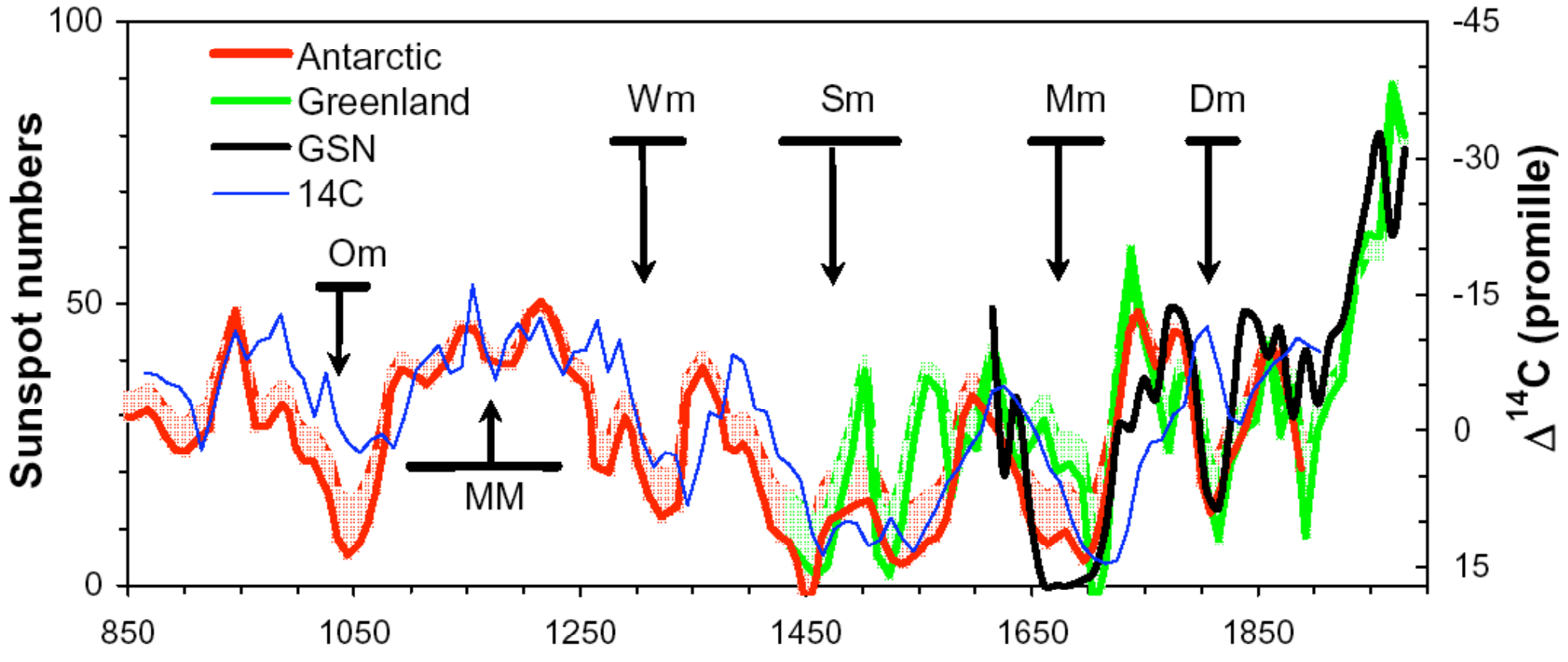

Figure 4. Time series of the sunspot number as reconstructed from $^{10}Be$ concentrations in ice cores from Antarctica (red) and Greenland (green). The corresponding profiles are bounded by the actual reconstruction results (upper envelope to shaded areas) and by the reconstructed values corrected at low values of the SN (solid curves) by taking into account the residual level of solar activity in the limit of vanishing SN. The thick black curve shows the observed group sunspot number since 1610 and the thin blue curve gives the (scaled) $^{14}C$ concentration in tree rings, corrected for the variation of the geomagnetic field. The horizontal bars with attached arrows indicate the times of great minima and maxima [J. Beer, S. Tobias, N.O. Weiss, *Solar Phys*. **181**, 237 (1998)]: Dalton minimum (Dm), Maunder minimum (Mm), Spörer minimum (Sm), Wolf minimum (Wm), Oort minimum (Om), and medieval maximum (MM). The temporal lag of $^{14}C$ with respect to the sunspot number is due to the long attenuation time for $^{14}C$ [E. Bard, G.M. Raisbeck, F. Yiou and J. Jouzel, *Earth Planet. Sci. Lett*. **150**, 453 (1997)]. Figure and modified caption from I. G Usoskin, et al., *Phys. Rev. Lett*. **91**, 211101-1 (2003). "promille" means parts per thousand.

The background given above shows that there is clearly a strong link between variations in solar activity and climate [13]. But how can the small, observed variations in solar irradiance have such a large impact on global temperature?

## Variations in cloud cover and cosmic rays

If solar variations are to play an important role in climate change, what is needed is a mechanism that enhances the effects of the relatively small variations in solar irradiance. The most likely candidate for this, although still controversial, is the modulation of cosmic ray flux by solar activity and the observed, correlated variations in the earth's albedo. However, cyclic variations in earth's climate following the sun's 11 year cycle cannot alone explain the warming over the last century. At most, such correlations show that the sun can affect climate, but for solar activity to be responsible for a significant portion of the last century's warming, there must be a centennial change. And, there is. Cosmic-ray intensity, as reconstructed from $^{10}Be$ concentrations in ice cores show a ~5-6% decrease over the twentieth century, corresponding to a 1% decrease in cloud cover.

The sun emits electromagnetic radiation and energetic particles known as the solar wind. A rise in solar activity affects the solar wind and the inter-planetary magnetic field by driving matter and magnetic flux trapped in the plasma of the local interplanetary medium outward, thereby creating what is called the heliosphere and partially shielding this volume, which includes the earth, from galactic cosmic rays—a term used to distinguish them from solar cosmic rays, which have much less energy. Long-term variations in the earth's magnetic field can also play a role. Solar variability not only affects the sun's irradiance, but also modulates incoming galactic cosmic radiation striking the earth's atmosphere. This is readily apparent in Figure 5 below [14]. Note the inverted scale for changes in solar irradiance.

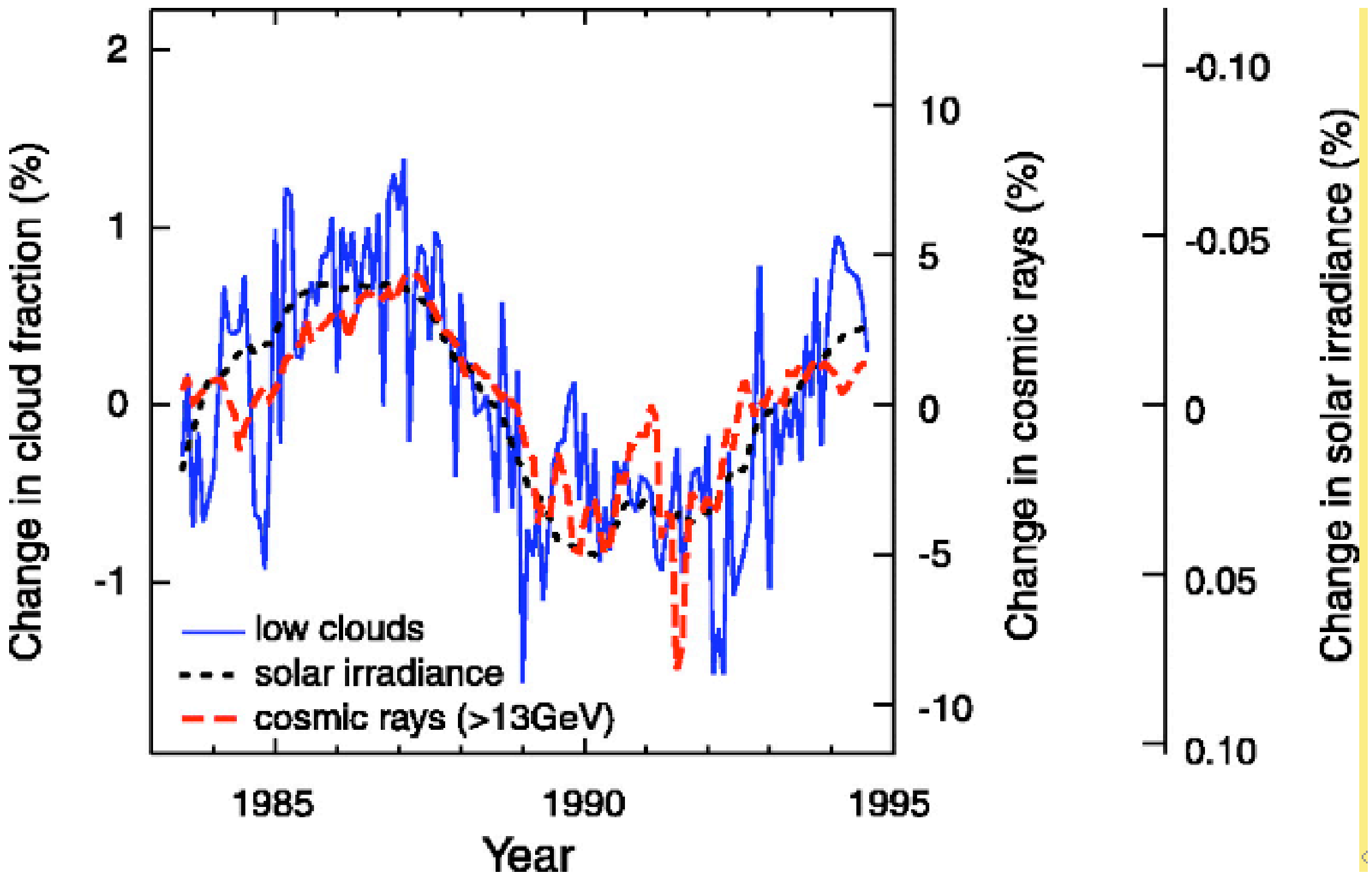

Figure 5. Variations of low-altitude cloud cover (less than about 3 km), cosmic rays, and total solar irradiance between 1984 and 1994. From K.S. Carslaw, R.G. Harrison, and J. Kirkby, *Science* **298**, 1732 (2002). Note the inverted scale for solar irradiance.

What Figure 5 shows is the very strong correlation between galactic cosmic rays, solar irradiance, and low cloud cover: When solar activity decreases, with a consequent small decrease in irradiance, the number of galactic cosmic rays entering the earth's atmosphere increases as does the amount of low cloud cover. This increase in cloud cover results in an increase in the earth's albedo, thereby lowering the average temperature. The sun's 11 year cycle is therefore not only associated with changes in irradiance, but also with changes in the solar wind, which in turn affect cloud cover by modulating the cosmic ray flux. This, it is argued, constitutes the strong positive feedback needed to explain the significant impact of small changes in solar activity on climate.

Cosmic ray fluxes can be reconstructed for the Phanerozoic period from iron meteorites. The following (controversial) figure [15] shows the strong correlation between this reconstructed flux and Phanerozoic temperature variations. The blue bands at the top of the figure correspond to an "icehouse" earth. The fitted curve in the bottom part of the figure is the calculated variation

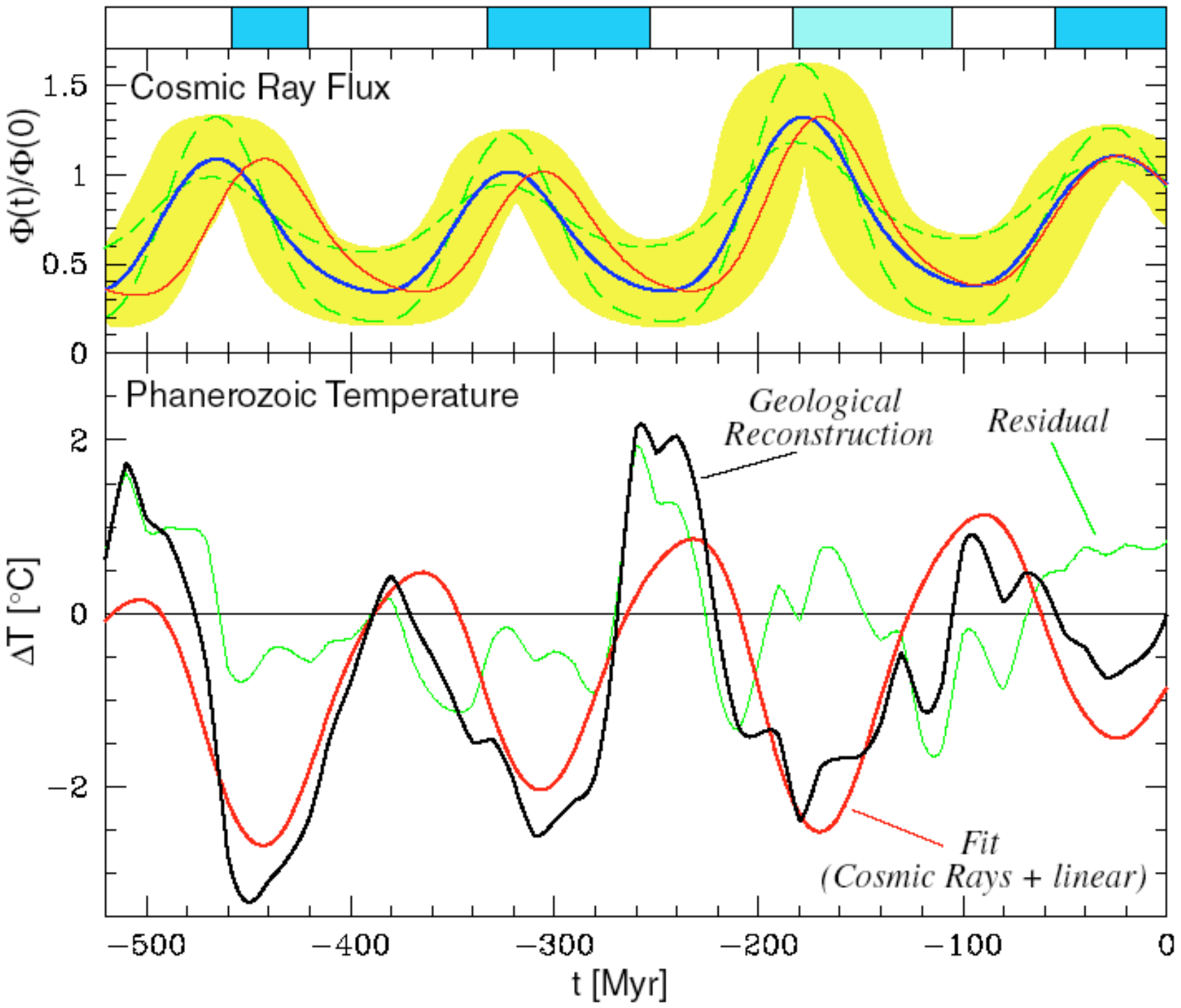

Figure 6. Cosmic ray flux (Φ) and tropical temperature anomaly (ΔT) variations over the Phanerozoic. From N.J. Shaviv and J. Veizer, *GSA Today* (July 2003).

in temperature (ΔT) due to the variation in galactic cosmic ray flux.

The correlations shown in Figures 5 and 6 clearly indicate that solar variations are an important driver of climate. What is missing is a robust explanation of the physical processes linking cosmic ray variations with cloud formation—cosmic rays, if they are to increase cloud cover, must increase the density of cloud condensation nuclei. But, the absence of an *accepted* model for this process does not negate the strong correlations shown in these figures. An overview of the physical processes involved has been given by Carslaw, et al. [3].

Finally, Figure 7 shows the centennial variation in cloud cover derived from a variety of indices [16].

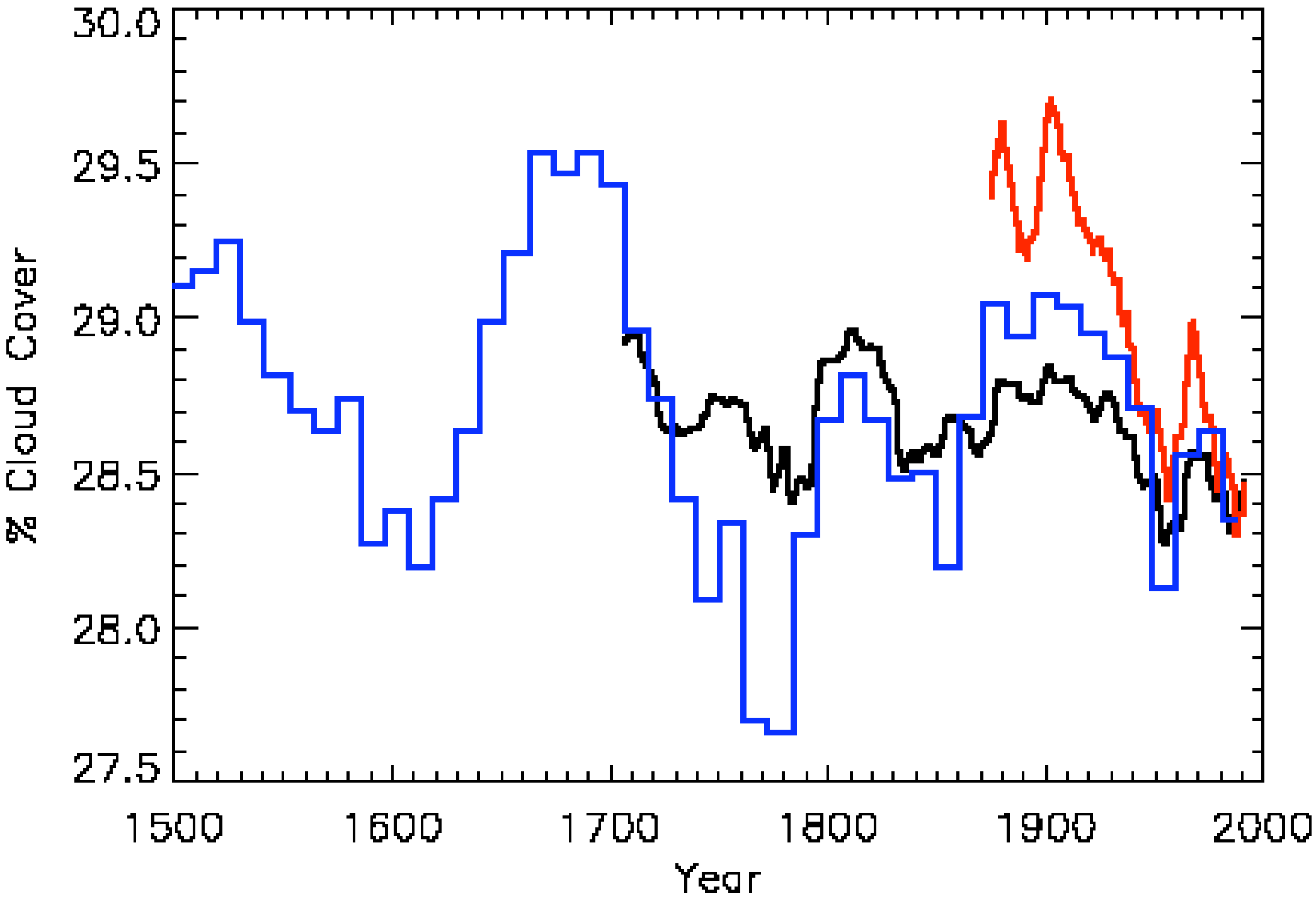

Figure 7. The 11-year smoothed reconstructed cloud cover for the whole earth derived from the Zurich Sunspot number (middle curve), the aa index (top curve), and the heliocentric potential (the curve extending from the year 1500). From E.P. Bago and C.J. Butler, *Astronomy and Geophysics* **41**, 18 (2000).

In Figure 7, the "heliocentric potential" is an electric potential centered on the sun, which is introduced to simplify calculations by substituting electrostatic repulsion for the interaction of cosmic rays with the solar wind. Its magnitude is such that the energy lost by cosmic rays in traversing this electric field to reach the earth's orbit is equal to the energy that would be lost by cosmic rays while interacting with the solar wind in passing through the solar system to reach the earth.

As can be seen from the figure, and as was mentioned earlier, there has been a decrease in low cloud cover by about 1% over the last century. This is consistent with the data in Figure 5 where a 1% decrease in cloud cover corresponds to a 5-6% decrease in galactic cosmic ray flux. Note also, the significant rise in cloud cover during the Little Ice Age (LIA) extending from the mid 17$^{th}$ century to the early 18$^{th}$.

**A phenomenological approach**

Two simple, phenomenological approaches to estimating the effect of solar variations on climate will be developed here. To do this we need the global temperature change during the LIA, which extended from about 1650 to 1710. While regional temperatures could have been much lower, estimates for the global change in temperature range from 0.3-0.5 $^{o}$C. The average value of 0.4 $^{o}$C will be used here.

As mentioned earlier, the data in Figure 1 were updated in 2004 and 2005. It is now believed [17] that the rise in solar irradiance since the LIA and 1996 was ~1 w/m$^2$, corresponding to the rescaling of 0.27 given in the draft 2007 IPCC report. Other work by Krivova, et al. [18], with data extending to the year 2000, finds the increase to be 1.3 w/m$^2$. The actual value of the total rise since the LIA does not affect the first approach used here. A rescaling of the data to a smaller overall rise in irradiance simply raises the sensitivity of the earth to changes in solar forcing while decreasing the forcing since 1900, the period of interest, so as to yield the same result as given below. Since it is easier to see the changes in irradiance from the LIA to the "quiet sun" period of around 1850, and the rise in irradiance from the later to the present day, Figure 1 will be used for calculational purposes.

Ignoring the Dalton minimum around 1810-1820, it can be seen from Figure 1 that the change in total solar irradiance between the LIA and about 1850, which is also a period of essentially constant carbon dioxide concentrations, is about 1.75 w/m$^2$. The value of 1.75 w/m$^2$ corresponds to a change in radiative forcing of $\delta F$ =0.3 w/m$^2$. The constancy of carbon dioxide concentrations over the period of 1600 to 1850 is important since it means that this gas was not a factor in the temperature changes associated with the LIA. It is generally accepted that these changes were due to variations in solar activity during this period.

The sensitivity of the earth to changes in solar forcing is then

$$\frac{\delta T}{\delta F} = \frac{0.4\ ^{o}\mathrm{C}}{0.3\ \mathrm{w\ m^{-2}}} = 1.33\ ^{o}\mathrm{C\ w^{-1}\ m^{2}}.$$

Since this sensitivity is due to changes in solar activity, it would include not only changes in radiative forcing due to changes in irradiance, but also increases in cloud cover due to increases in galactic cosmic rays during the LIA, should they have been a factor. Note that this value of climate sensitivity is comparable to that given by the data point designated as (20) in Figure 8(A) below.

Now from Figure 1, the change in solar irradiance since 1900, a time during which carbon dioxide concentrations were also rising, is about 1.5 w/m$^2$. This implies a radiative forcing of $\Delta F$ = 0.26 w/m$^2$, so the change in temperature due to solar variation (including any change in albedo due to a change in cosmic ray flux) is then

$$\Delta T = \Delta F\,\frac{\delta T}{\delta F} = 1.33\ ^{o}\mathrm{C\ w^{-1}\ m^{2}}\ \mathrm{X}\ 0.26\ \mathrm{w\ m^{-2}} = 0.35\ ^{o}\mathrm{C}.$$

The rise in temperature over the last century has been about 0.7 $^{o}$C, so this simple, phenomenological approach leads to the conclusion that the increase in solar activity over this period is responsible for 50% of the rise. Again, the rescaling of the data in Figure 1 by later work will not change this temperature rise since the product of the sensitivity and forcing is invariant under such a rescaling.

A second, independent calculation can be made for the temperature rise due to increased solar activity by using the decrease in cloud cover over the last century. To do this one must revisit the issue of overall climate sensitivity mentioned in the second section of this paper. The following figure shows the data for different empirical estimates of climate sensitivity in the case where cosmic ray flux is included and also when it is not.

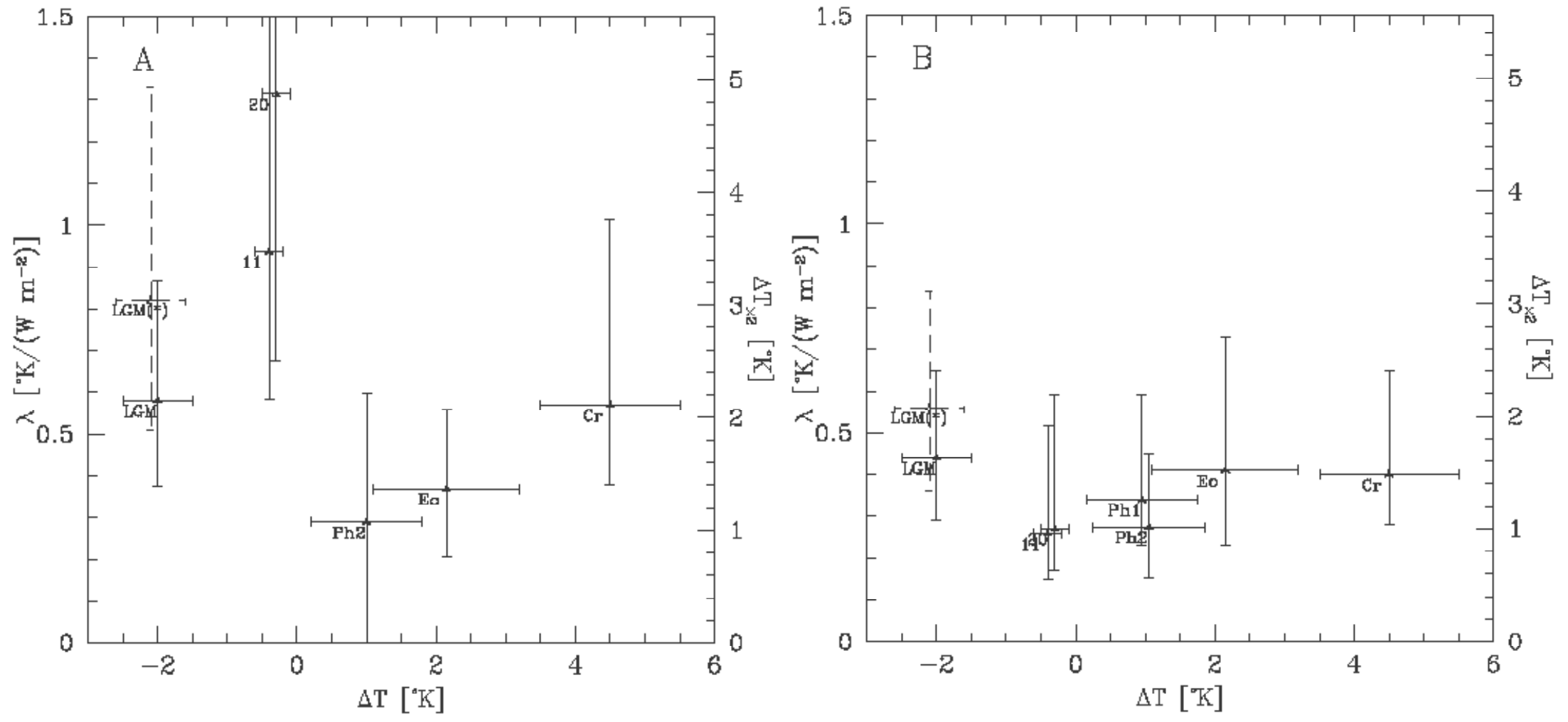

Figure 8. Estimated sensitivity $\lambda$ as a function of the difference between the temperature today and the time in the past for which the sensitivity was calculated. Values are for the Last Glacial Maximum (LGM), the 11 year solar cycle over the past 200 years (11), $20^{th}$ century global warming (20), Phanerozoic by comparison of the tropical temperature to cosmic ray flux variations (Ph1) or carbon dioxide variations (Ph2), Eocene (Eo) and Mid-Cretaceous (Cr). From N. J. Shaviv, *Journal of Geophysical Research* **110**, A08105 (2005).

The average of the different empirical estimates of climate sensitivity in Figure 8(A), where the link between cosmic ray flux and climate is ignored, is $\lambda = 0.54\ ^oC\ w^{-1}\ m^2$, very close to the value mentioned in Section 2 of this paper and generally accepted by the community. If, on the other hand, this link is included the average of the data in Figure 8(B) is $\lambda = 0.34\ ^oC\ w^{-1}\ m^2$. What is interesting here is that the scatter in the data is much less if the cosmic-ray link to climate is included. This validates the implicit assumption in the work cited that climate sensitivity to changes in cosmic ray flux remains constant at different times and under varying climate conditions. The latter value of climate sensitivity will be used to make another independent estimate of the fraction of warming since 1900 due to changes in solar activity.

By considering the change in the earth's albedo due to a change in cloud cover, it was shown in section two of this paper that a 1% decrease in total cloud cover—which is what has been observed since 1900—corresponds to a net radiative forcing of 0.62 w/m$^2$. Using the value of climate sensitivity that includes the climate link to cosmic ray flux, this forcing implies a change in earth's temperature since 1900, due to the change in albedo resulting from a reduction in cloudiness, of

$$\Delta T_{CL} = 0.34\ ^oC\ w^{-1}\ m^2\ \ X\ 0.62\ w\ m^{-2} = 0.21\ ^oC.$$

To this must be added the temperature rise, $\Delta T_{IR}$, due to the increase in irradiance over the same period. In the discussion of the LIA, the radiative forcing due to the rise in irradiance since 1900 was given as 0.26 w/m$^2$. Interestingly enough, a rescaling of the data in Figure 1 will affect this calculation, but the radiative forcing obtained is less than that given by the IPCC in its 2007 Summary for Policymakers. They give the rise in radiative forcing since 1750 due to the increase in solar irradiance as 0.12 w/m$^2$. As can be seen from Figure 1, total solar irradiance in 1750 and 1900 were essentially the same. The value given by the IPCC will be used here. The temperature to be added is then

$$\Delta T_{IR} = 0.34\ ^oC\ w^{-1}\ m^2\ \ X\ 0.12\ w\ m^{-2} = 0.04\ ^oC.$$

Therefore, the total rise in the earth's temperature since 1900 due to changes in cloud cover and irradiance is $\Delta T_{CL} + \Delta T_{IR}$ = 0.25 °C, or 36% of the observed temperature rise since 1900. The two approaches imply that solar activity is responsible for ~36-50% of the warming during the last century. This fraction is comparable to other estimates found in the literature [19, 20, 21]. Notice that using the higher value of climate sensitivity, where the link between galactic cosmic ray flux and climate is ignored, and which is closer to the generally accepted value, would lead to a greater fraction of the warming since 1900 being due to increased solar activity.

One consequence of 36-50% of the warming since 1900 being due to increased solar activity is that it implies that the IPCC is using too high a value for $\alpha$ in their simplified expression relating radiative forcing to changes in carbon dioxide concentration. The IPCC in its 2007 Summary for Policy Makers maintains that the total net anthropogenic radiative forcing since 1750 is 1.6 w/m$^2$, which is essentially the same as the anthropogenic radiative forcing since 1850. Since the other factors in the IPCC table of radiative forcing components essentially cancel (see Table 1), all of this forcing is due to the increase in carbon dioxide concentration.

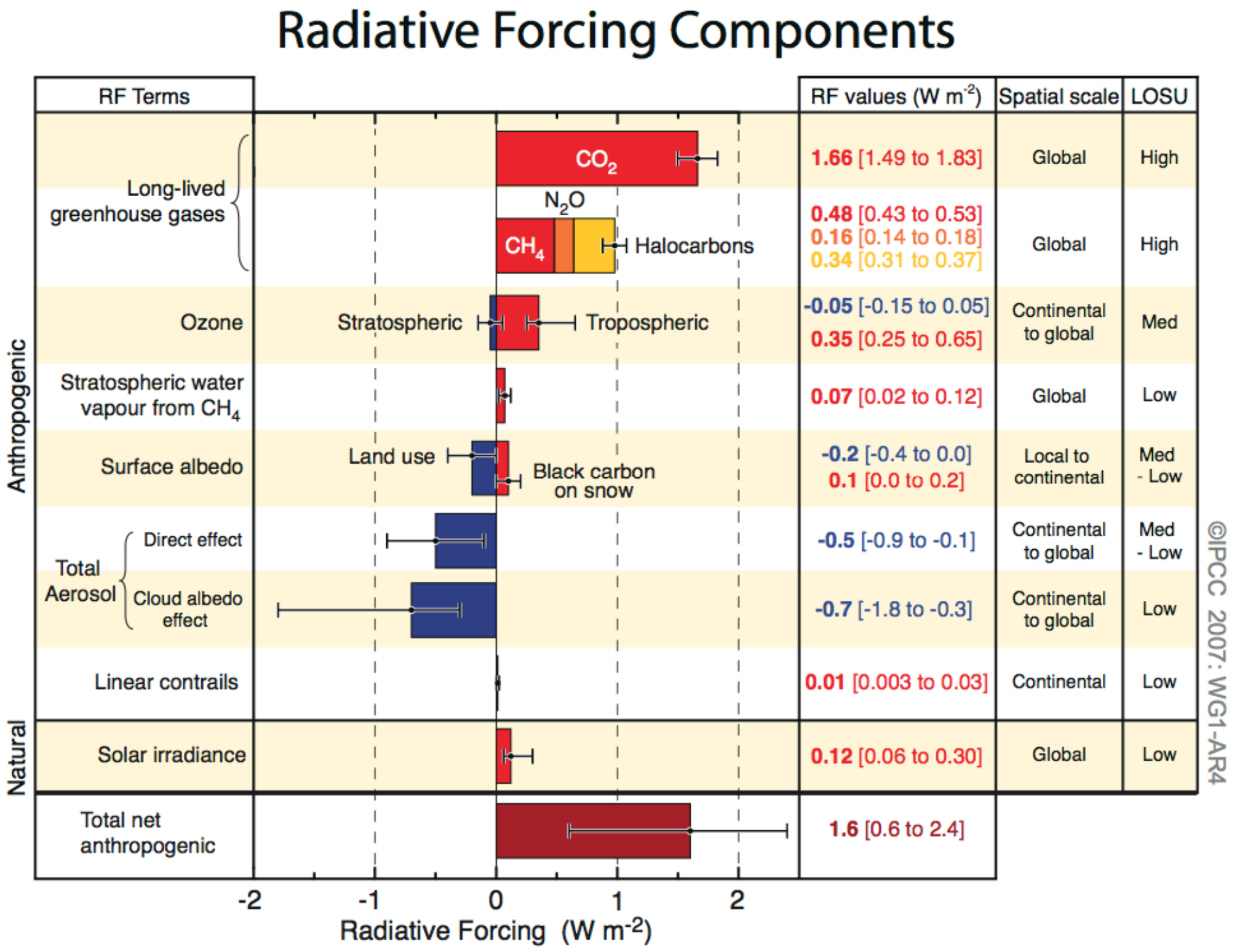

Table 1. Radiative forcings since 1750. From: Climate Change 2007, The Physical Basis, Summary for Policymakers, Figure SPM-2.

The corresponding rise in temperature since 1750, using the IPCC net anthropogenic forcing of 1.6 w/m$^2$ and the generally accepted overall value of climate sensitivity of 0.53 °Cw$^{-1}$m$^2$, is 0.85 °C.

The global temperature change since the period 1850-1900 is about 0.70 °C. Subtracting 0.3 °C, the average of the two values found above as being due to increased solar activity, leaves 0.4 °C as resulting from the increase in carbon dioxide concentration. Using the IPCC methodology, 0.4 °C corresponds to a radiative forcing of 0.84 w/m$^2$. Therefore, the coefficient $\alpha$ that should be used is

$$\alpha = \frac{\Delta F}{\ln\left(\frac{C}{C_0}\right)} = \frac{0.84 \text{ w m}^{-2}}{\ln\left(\frac{381}{300}\right)} = 3.5 \text{ w m}^{-2}.$$

This value is significantly lower than those given in the introduction.

**Conclusion**

Some of the arguments and data behind the contention that the earth's climate could be affected by changes in cloud cover caused by variations in the galactic cosmic ray flux have been briefly summarized. These data strongly imply that a relation between cosmic rays intensity and cloud cover may explain how relatively small changes in solar activity can produce much larger changes in the earth's climate. While the correlation is robust, there is still no generally accepted mechanism, although a number have been proposed. This is not surprising since the microphysical processes in clouds are quite complex and this is an ongoing area of research. Nevertheless, a lack of an accepted microphysical process for increasing cloud condensation nuclei does not justify minimizing the impact of solar variations on climate.

A simple, phenomenological approach was used to obtain two estimates of the fraction of the global temperature rise since 1900 due to a rise in solar activity. These methods yield a value of 36-50% for the solar fraction. This value is consistent with other estimates in the literature.

According to the 2007 IPCC Summary for Policymakers, essentially all of the anthropogenic radiative forcing since 1750 is due to the increase in carbon dioxide concentration. This means that if the sun is responsible for 36-50% of the temperature rise since 1900, the IPCC is using too large a value for its coefficient relating radiative forcing to changes in carbon dioxide concentration. This is important because a smaller value of $\alpha$ reduces the sensitivity of the earth's climate to increases in carbon dioxide concentrations—a result that has significant policy implications.